\documentclass{aa}

\usepackage{graphicx}
\usepackage{txfonts}

\newcommand{\C}{\v{C}erenkov }  

\begin{document}

\title{An unidentified TeV source in the vicinity of Cygnus OB2}

\titlerunning{A TeV source in the vicinity of Cygnus OB2}

\author{
F.~Aharonian\inst{1},
A.~Akhperjanian\inst{7},
M.~Beilicke\inst{4},
K.~Bernl\"ohr\inst{1},
H.~B\"orst\inst{5},
H.~Bojahr\inst{6},
O.~Bolz\inst{1},
T.~Coarasa\inst{2},
J.~Contreras\inst{3},
J.~Cortina\inst{2},
S.~Denninghoff\inst{2},
V.~Fonseca\inst{3},
M.~Girma\inst{1}
N.~G\"otting\inst{4},
G.~Heinzelmann\inst{4},
G.~Hermann\inst{1},
A.~Heusler\inst{1},
W.~Hofmann\inst{1},
D.~Horns\inst{1},
I.~Jung\inst{1},
R.~Kankanyan\inst{1},
M.~Kestel\inst{2},
J.~Kettler\inst{1},
A.~Kohnle\inst{1},
A.~Konopelko\inst{1},
H.~Kornmeyer\inst{2},
D.~Kranich\inst{2},
H.~Krawczynski\inst{9},
H.~Lampeitl\inst{4},
M.~Lopez\inst{3},
E.~Lorenz\inst{2},
F.~Lucarelli\inst{3},
N.~Magnussen\inst{10},
O.~Mang\inst{5},
H.~Meyer\inst{6},
M.~Milite\inst{4},
R.~Mirzoyan\inst{2},
A.~Moralejo\inst{3},
E.~Ona\inst{3},
M.~Panter\inst{1},
A.~Plyasheshnikov\inst{1,8},
J.~Prahl\inst{4},
G.~P\"uhlhofer\inst{1},
G.~Rauterberg\inst{5},
R.~Reyes\inst{2},
W.~Rhode\inst{6},
J.~Ripken\inst{4},
A.~R\"ohring\inst{4},
G.~P.~Rowell\inst{1},
V.~Sahakian\inst{7},
M.~Samorski\inst{5},
M.~Schilling\inst{5},
F.~Schr\"oder\inst{6},
M.~Siems\inst{5},
D.~Sobzynska\inst{2},
W.~Stamm\inst{5},
M.~Tluczykont\inst{4},
H.J.~V\"olk\inst{1},
C.~A.~Wiedner\inst{1},
W.~Wittek\inst{2} (HEGRA Collaboration),
Y.~Uchiyama\inst{11}, 
T.~Takahashi\inst{11}}

\institute{Max-Planck-Institut f\"ur Kernphysik, Postfach 103980, D-69029 Heidelberg, Germany
\and Max-Planck-Institut f\"ur Physik, F\"ohringer Ring 6, D-80805 M\"unchen, Germany
\and Universidad Complutense, Facultad de Ciencias F\'{\i}sicas, Ciudad Universitaria, E-28040 Madrid, Spain
\and Universit\"at Hamburg, Institut f\"ur Experimentalphysik, Luruper Chaussee 149, D-22761 Hamburg, Germany
\and Universit\"at Kiel, Institut f\"ur Experimentelle und Angewandte Physik, Leibnizstra{\ss}e 15-19, D-24118 Kiel, Germany
\and Universit\"at Wuppertal, Fachbereich Physik, Gau{\ss}str.20, D-42097 Wuppertal, Germany
\and Yerevan Physics Institute, Alikhanian Br. 2, 375036 Yerevan, Armenia
\and On leave from Altai State University, Dimitrov Street 66, 656099 Barnaul, Russia
\and Now at Washington University, St. Louis MO 63130, USA
\and Now at IFAE, Unversitat Aut\`onoma de Barcelona, Spain
\and Institute of Space and Astronautical Science, 3-1-1 Yoshinodai, Sagamihara, Kanagawa 229-8510, Japan
}

\authorrunning{Aharonian et al.}

\date{Received / Accepted}

\offprints{G.P. Rowell, D. Horns\\
\email{\scriptsize Gavin.Rowell@mpi-hd.mpg.de,Dieter.Horns@mpi-hd.mpg.de}}

\abstract{
Deep observation ($\sim113$ hrs) of the Cygnus region at TeV energies using the HEGRA
stereoscopic system of air \C telescopes has serendipitously revealed a
signal positionally inside the core of the OB association Cygnus OB2, at the edge of 
the 95\% error circle of the EGRET source 3EG J2033+4118, and $\sim0.5^\circ$ north of Cyg X-3.
The source centre of gravity is 
RA $\alpha_{\rm J2000}$: 20$^{\rm hr}$ 32$^{\rm m}$ 07$^{\rm s}$ $\pm 9.2^{\rm s}_{\rm stat} 
\pm2.2^{\rm s}_{\rm sys}$,
Dec $\delta_{\rm J2000}$: +41$^\circ$  30$^\prime$  30$^{\prime\prime}$ $\pm 2.0^\prime_{\rm stat}
\pm 0.4^\prime_{\rm sys}$.
The source is steady, has a post-trial significance of +4.6$\sigma$, indication for extension with radius
$5.6^\prime$ at the $\sim3\sigma$ level, and has a differential power-law flux with
hard photon index of $-1.9 \pm0.3_{\rm stat}\pm0.3_{\rm sys}$. The integral flux above 1 TeV amounts $\sim$3\% 
that of the Crab.
No counterpart  for the TeV source at other wavelengths is presently identified,
and its extension would disfavour an exclusive pulsar or AGN origin. 
If associated with Cygnus OB2, this dense concentration of young, massive stars
provides an environment conducive to multi-TeV particle acceleration and likely subsequent interaction with a nearby
gas cloud. Alternatively, one could envisage $\gamma$-ray production via a jet-driven termination shock.
 
}

\maketitle

%

\section{Introduction}
\label{introduction}

The current generation of ground-based imaging atmospheric \C telescopes offer coverage of the 
multi GeV to TeV $\gamma$-ray sky at 
centi-Crab sensitivity and arc-minute resolution. 
Stereoscopy employed by the HEGRA CT-System at La Palma (Daum et al. \cite{Daum:1}) 
offers highly accurate reconstruction of event directions at angles up to $\sim3^\circ$ off-axis. 
Results here are taken from data originally devoted to Cyg X-3, and the EGRET source
GeV~J2035+4214 (Lamb \& Macomb \cite{Lamb:1}). The separation between
these objects 
($\sim1.5^\circ$) permits a combined analysis given the overlap in their CT-System fields of view (FOV).
This letter presents analysis details 
and observational properties of a serendipitously discovered TeV source in these data.
A brief discussion concerning astrophysical origin and location of this
new source is also presented.


\section{Data Analysis and Results}
\label{observations}

The HEGRA system of imaging atmospheric \C telescopes (IACT-System), consists of 5 identical 
telescopes operating in coincidence for the
stereoscopic detection of air showers induced by primary $\gamma$-rays
in the atmosphere.
In data dedicated to Cyg X-3, alternate $\sim$20 minute runs
targeting the Cyg X-3 position $\pm 0.5^\circ$ in declination were
taken during moonless nights of Aug-Sept 1999, Sept-Oct 2000 and Jun-Oct 2001.
Likewise in data dedicated to GeV J2035+4124, $\sim$20 minute runs were obtained tracking directly
the GeV source during Jul-Aug 2001. In total, three
tracking positions are present in combined data.
After quality checks, a total of 112.9 hours data are available for analysis.
Preferential selection of $\gamma$-ray-like events (against
the cosmic-ray background) 
is achieved by using the difference between the reconstructed 
and assumed event direction, $\theta$,
and the mean-scaled-width parameter, 
$\bar{w}$ (Konopelko \cite{Konopelko:2}).
In searching for weak point-like and marginally extended sources,
so-called tight cuts are considered optimal given the angular resolution of the CT-System ($<0.1^\circ$):
$\theta < 0.12^\circ$ and $\bar{w} < 1.1$, where we use algorithm '3' as described by Hofmann et al. (\cite{Hofmann:1})
for the event direction reconstruction. The number
of images per event, $n_{\rm tel}$, used for calculating $\theta$ and
$\bar{w}$ was also {\em a priori} chosen at $n_{\rm tel}\ge3$.
Monte Carlo simulations (Konopelko et al. \cite{Konopelko:1}) and
tests on real sources have shown that $n_{\rm tel}=2$ events
contribute little to the overall sensitivity.

\subsection{Source Search and Background Estimates}
\begin{figure}
\centering
\includegraphics[width=8.5cm]{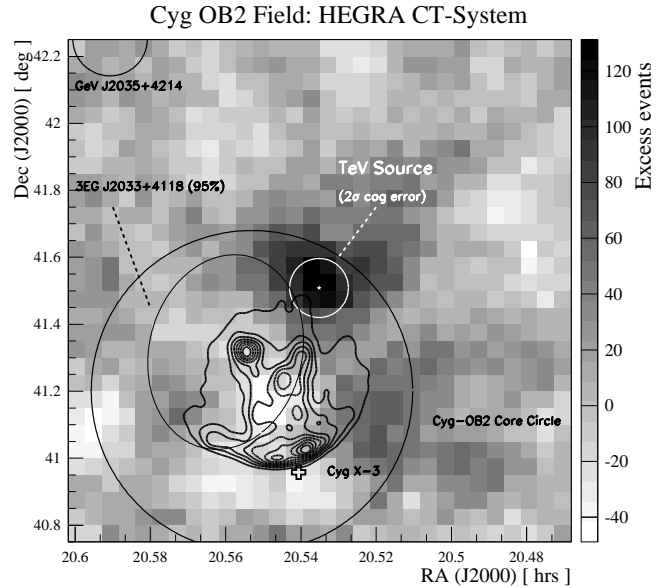}
\caption[]{Skymap (1.5$^\circ \times 1.5^\circ$ view at 0.05$^\circ\times0.05^\circ$ binning)
of excess events $s-\alpha b$, using the template background model. At each bin, the excess is
estimated from events within a radius
$\theta=0.12^\circ$. Included are 95\% error ellipses
           of various EGRET sources, the core of Cygnus OB2
           (Kn\"odlseder \cite{Knodlseder:1}), the TeV COG (star), its 2$\sigma$ error circle,
           and the location of Cyg X-3. ASCA GIS contours (2-10 keV) are overlayed.}
\label{fig:skymap}
\end{figure}

In searching for new TeV sources, {\em skymaps} of event
direction excesses over the RA \& Dec plane are generated after having estimated the 
background over the FOV. A new empirically-based {\em template} background model has
been developed with the goal of simple generation of skymaps.
The template background comprises events normally rejected according 
to the $\bar{w}$ criterion. We define the number of events in the $\gamma$-ray regime $s$ from $\bar{w}<1.1$, 
and for the template background $b$ from $1.3<\bar{w}<1.5$. A necessary correction applied
to the template background accounts for differences in radial
profile between the two $\bar{w}$ regimes.
A normalisation $\alpha$, to derive excess events $s-\alpha b$ at
some position in the FOV, accounts for differences 
in the total number of events in the two $\bar{w}$ regimes. 
A full description of the template model appears in Rowell \cite{Rowell:2}.
Fig.~\ref{fig:skymap} presents the resulting excess skymap.
The template model was used in discovering the TeV source which is evident $\sim 0.5^\circ$ north of 
Cyg X-3. An event-by-event
centre of gravity (COG) calculation (Table~\ref{tab:numbers}a),
weighting events with $\pm$1 from 
the $s$ and $\alpha b$ regimes
respectively is performed. The COG accuracy is limited by a systematic pointing error of $\sim25^{\prime\prime}$
(P\"uhlhofer et al. \cite{Puehl:1}).
A pre-trial significance at the COG position of +5.9$\sigma$ is obtained, summing 
events within $\theta=0.12^\circ$ 
(Table~\ref{tab:numbers}b). Statistical trial factors arise from the initial ´discovery´
skymap (different to that in Fig.~\ref{fig:skymap}) in which event directions are independently
summed in 1100 bins of size 0.1$^\circ \times 0.1^\circ$.
Assuming 1100 trials are accrued in locating the COG, the
post-trial probability $P_t=1.0-(1-P)^{1100}$ for $P$ the pre-trial
probability  (one-sided $P=1.9\times10^{-9}$, or +5.9$\sigma$), is then calculated as
$P_t=2.1\times10^{-6}$. This gives a post-trial significance of
+4.6$\sigma$. 1100 is actually a slightly conservative trial estimate since oversampling of the 
$\gamma$-ray point spread function (PSF) by a factor $\sim1.5$
occurs in the discovery skymap.
To verify results using the template model, we make use of
a conventional type of background model employing background regions
displaced from the on-source region spatially in the FOV but derived
from the same $\bar{w}<1.1$ regime. Background events are taken from ring-segments with
matching trigger characteristics to that of the source region.
A normalisation $\alpha$ according to the solid angle ratio between background and on-source regions is then applied. 
Results using this so-called {\em ring} model (Table.~\ref{tab:numbers}b) are consistent with those from the template model.
\begin{table}[ht]
  {\centering
  {\footnotesize
  \begin{tabular}{l}
   \fbox{\bf (a) Centre of Gravity}\\
  \end{tabular}
  \begin{tabular}{lll}
      RA $\alpha_{2000}$:   &  20$^{\rm hr}$ 32$^{\rm m}$ 07$^{\rm s}$ & $\pm 9.3^{\rm s}_{\rm stat}$ $\pm 2.2^{\rm s}_{\rm sys}$ \\
      Dec $\delta_{2000}$:  &  41$^\circ$   \,\,30$^\prime$  \,\,30$^{\prime\prime}$ & $\pm 2.0^\prime_{\rm stat}$ $\pm 0.4^\prime_{\rm sys}$ \\
  \end{tabular}\\
  \begin{tabular}{l}\\
   \fbox{\bf (b) Tight cuts: $\theta<0.12^\circ$, $\bar{w}<1.1$, $n_{\rm tel} \geq 3$} \\
  \end{tabular}
  \begin{tabular}{lccccc} 
  Background        &  $s$ & $b$   & $\alpha$ & $s-\alpha\,b$ & $S$   \\ \hline
  Template          &  523 & 2327  & 0.167    &   134         & {\bf +5.9} \\
  Ring              &  523 & 4452  & 0.089    &   128         & {\bf +5.9} \\ \hline
  \end{tabular} \\
  \begin{tabular}{l} \\
  \fbox{\bf(c) Tight cuts on tracking subsets}\\
  \end{tabular}
  \begin{tabular}{lcccccc}
  Back.          & $t$ & $\eta$ &  $s$ & $b$   & $\alpha$ & $S$   \\ \hline
  \multicolumn{7}{c}{  --- Cyg X-3 $\delta-0.5^\circ$ ---}\\
  Template                      & 39.5 &  0.69 & 148  &  647   & 0.172 & {\bf +3.0} \\
  Ring                          &      &       & 148  & 1994   & 0.057 & {\bf +3.3} \\
  \multicolumn{7}{c}{  --- Cyg X-3 $\delta$+0.5$^\circ$ ---}\\
  Template                      & 45.4 &  1.00 & 276  & 1214  & 0.170  & {\bf +4.2} \\
  Ring                          &      &       & 276  & 1266  & 0.168  & {\bf +3.8} \\
  \multicolumn{7}{c}{  --- GeV J2035 ---}\\
  Template                      & 28.0 &  0.68 &  99  &  472  & 0.156  & {\bf +2.6} \\ 
  Ring                          &      &       &  99  & 1193  & 0.057  & {\bf +3.4} \\ \hline
  \multicolumn{7}{l}{\scriptsize $t$: Observation time (hrs)}\\
  \multicolumn{7}{l}{\scriptsize $\eta$: Estimated $\gamma$-ray trigger effic. {\em cf.} on-axis.}
  \end{tabular}\\
  \begin{tabular}{l} \\
  \fbox{\bf (d) Tight cuts on $n_{\rm tel}$ subsets}\\
  \end{tabular}
  \begin{tabular}{lccccc} 
  Back.           &  $s$ & $b$   & $\alpha$ & $s-\alpha b$ &     $S$   \\ \hline
  \multicolumn{6}{c}{  --- $n_{\rm tel}=2$ ---}\\
  Template        & 387  &  865  & 0.433  & 12 & {\bf +0.8} \\
  Ring            & 387  & 4619  & 0.082  &  8 & {\bf +0.5} \\
  \multicolumn{6}{c}{  --- $n_{\rm tel}=3$ ---}\\
  Template        & 272  &  904  & 0.224  & 70  & {\bf +4.1} \\
  Ring            & 272  & 2691  & 0.086  & 41  & {\bf +2.6} \\
  \multicolumn{6}{c}{  --- $n_{\rm tel}=4$ ---}\\
  Template        & 133  &  774  & 0.130  & 32  & {\bf +2.9} \\ 
  Ring            & 133  &  1110 & 0.088  & 44  & {\bf +3.2} \\
  \multicolumn{6}{c}{  --- $n_{\rm tel}=5$ ---}\\
  Template        & 118  & 777   &  0.089 & 50  & {\bf +5.1} \\
  Ring            & 118  & 651   &  0.102 & 52  & {\bf +5.4} \\ \hline
  \end{tabular}\\
  \begin{tabular}{l} \\
   \fbox{\bf (e) Spectral Cuts$^\dagger$: $\theta<0.224^\circ$, $\bar{w}<1.1$, $n_{\rm tel} \geq 3$}\\
  \end{tabular}
  \begin{tabular}{lccccc}
  Back.       &  $s$ & $b$   & $\alpha$ & $s-\alpha\,b$ & $S$   \\ \hline
  Ring        &  366 & 3222  & 0.087   &   86         & {\bf +4.7} \\ \hline
  \multicolumn{6}{l}{\scriptsize $^\dagger$: Aharonian et al. \cite{Aharonian:2} summarise other spectral cuts.} \\ 
  \end{tabular}\\
  }
  }
  \caption[]{Summary of numerical results for the TeV source,
             under two background models. Here, $s$ and $b$ are the 
             resulting event numbers for the $\gamma$-ray-like and background $\bar{w}$ regimes respectively, and
             $s-\alpha b$ is the derived excess using a normalisation $\alpha$. $S$ denotes the
             excess significance using Eq. 17 of Li \& Ma \cite{Li:1}. 
             See section~\ref{observations} for definitions of $\theta$ and $\bar{w}$.}
  \label{tab:numbers}
\end{table}

\subsection{Observational Properties of the TeV Source}
\begin{figure}[t]
\centering
\includegraphics[width=8.0cm]{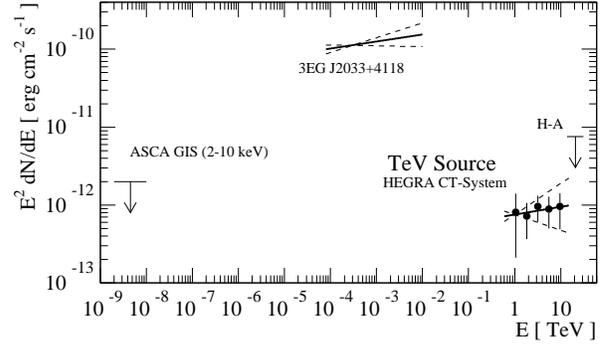}
\caption{\label{fig:spectrum}Differential energy fluxes of the TeV source and other results. 
         'H-A' is the AIROBICC 90\% confidence level upper limit
          (Prahl \cite{Prahl:1}) at the TeV COG converted 
         to differential form at 20.8 TeV assuming a spectral photon index of $-$2.0. 
         We interpret the 3EG J2033+4118 flux as an upper limit.
         The ASCA GIS 99\% upper limit assumes a photon index of 2, and $N_H=10^{22}$ cm$^{-2}$. 
         Data fits (solid lines) and their 1$\sigma$ statistical errors in photon index (dashed lines) are shown. 
         TeV data are well fit ($\chi^2_\nu=0.23/3$) by a power law (Eq.~\ref{eq:spectrum}).}
\end{figure}

Splitting data firstly according to 
their three tracking positions reveals commensurate source contributions 
(Table~\ref{tab:numbers}c). The source is also found
to develop linearly with the cumulative number of background events.
Such tests suggest consistency with a steady source during the three years of data
collection. We have also verified that after cuts a constant background acceptance 
throughout the dataset is observed and that the event excess in $\bar{w}$-space appears consistent
with that of a true $\gamma$-ray population. 
\begin{figure}
\centering
\includegraphics[width=7.5cm]{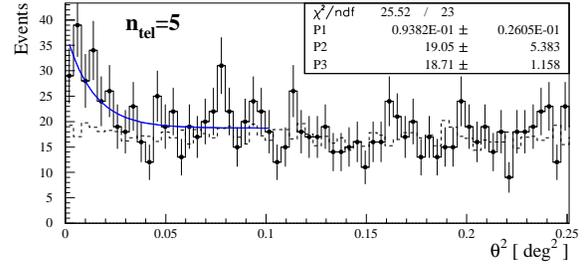}
\caption[]{Distribution of $\theta^2$ for $\theta$ calculated from the
           COG (solid histogram \& filled dots), against a
           background derived from the template model
           (dashed hist). The convolved radial Gaussian fit $F=P3 + P2 \exp(-\theta^2/(P1^2+\sigma^2_{\rm pt}))$ is
 	   indicated by the solid line with $P1=\sigma_{\rm src}=$0.094$^\circ \pm0.026^\circ$ the
intrinsic source size. The PSF width $\sigma_{\rm pt}$=0.070$^\circ$ is estimated from Crab data.}
\label{fig:quality}
\end{figure}
To determine source size, we fit a radial Gaussian convolved with the point spread
function (determined from Crab data) to the
excess events as a function of $\theta^2$, using a subset of events with 
the best angular resolution ($n_{\rm tel}=5$) for which errors
are minimised (Fig.~\ref{fig:quality}). 
The intrinsic size of the TeV source is estimated at $\sigma_{\rm src}=5.6^\prime\pm1.7^\prime$.
Correlations between the fit parameters suggest that the significance for a non-zero source size
is at the $\sim3.0\sigma$ level rather than the 3.5$\sigma$ level indicated above.
A breakdown of the excess with $n_{\rm
tel}$ also shows that the $n_{\rm tel}$=5 exclusive subset 
contributes strongly to the excess
(Table~\ref{tab:numbers}d). Such behaviour is
suggestive of a generally hard spectral index given that higher trigger
multiplicities are favoured by higher energy events. 
For the energy spectrum calculation and selection cuts, we follow
the method of Aharonian et al. (\cite{Aharonian:2}) using
effective collection areas appropriate for on-axis and
$\sim1^\circ$ off-axis sources as per the 
exposure efficiency for the TeV source in these data.
A tight cut $\bar{w}<$1.1, as opposed to the
less-restrictive $\bar{w}<1.2$ is also used. For energies below $\sim$0.8 TeV the effective collecting area 
decreases markedly sharper at positions beyond $\sim1.0^\circ$ off-axis compared to positions nearer on-axis.
Limiting our fit therefore to energies $>$0.8 TeV reduces systematic
errors. A so-called loose cut $\theta<0.224^\circ$
using the ring background model is used in deriving the energy bin-by-bin excess since a loose cut 
in $\theta$ improves the $\gamma$-ray selection efficiency according
to the moderately-extended nature of the source. 
Results are shown in Table~\ref{tab:numbers}e and Fig.~\ref{fig:spectrum}, with the spectrum being well fit
by a pure power law with generally hard photon
index. Systematic errors are estimated from changes in bin centres and uncertainties in Monte Carlo-derived
collection areas:
\begin{eqnarray}
 dN/dE  & = & N\,\, (E/1\, {\rm TeV})^\gamma \,\,\,\, {\rm ph\, cm^{-2}\, s^{-1}\, TeV^{-1}} \label{eq:spectrum} \\
 N      & = & \,4.7\, (\pm2.1_{\rm stat} \pm1.3_{\rm sys})\, \times 10^{-13} \nonumber \\
 \gamma & = & -1.9 \, (\pm0.3_{\rm stat} \pm0.3_{\rm sys})  \nonumber 
\end{eqnarray}
The integral flux at $F(E>1$ TeV) = $4.5\, (\pm1.3_{\rm stat})\times 10^{-13}$ ph cm$^{-2}$ s$^{-1}$ amounts to 
2.6\% that of the Crab.

\section{Discussion \& Conclusion}

The OB association Cygnus OB2 is unique for its compact nature 
and extreme number of member OB and O stars (eg. Kn\"odlseder
\cite{Knodlseder:1}), and both theoretical and observational
grounds for non-thermal particle acceleration have long-been discussed 
(eg. Montmerle \cite{Montmerle:1}, Cass\'{e} \& Paul \cite{Casse:1}, V\"olk \& Forman \cite{Volk:1}, 
White \& Chen \cite{White:1}). 
The TeV source is positioned inside the core of Cygnus OB2 as defined
by Kn\"odlseder (\cite{Knodlseder:1}). 
Assuming the TeV source is as distant as Cygnus~OB2 (1.7 kpc), a luminosity $\sim 10^{32}$ erg
s$^{-1}$ above 1 TeV is implied, well within the
kinetic energy (KE) budget of Cygnus OB2 estimated recently by Lozinskaya et al. (\cite{Lozinskaya:1}) at 
a few$\times 10^{39}$ erg s$^{-1}$, and also within the KE budget of a number of notable member stars
(eg. Massey \& Thompson \cite{Massey:1}, Manchanda et al. \cite{Manchanda:1}, Benaglia et al. \cite{Benaglia:1}).
So far no counterparts at other wavelengths are identified. No massive or 
luminous Cygnus OB2 star of note discussed recently (eg. Massey \& Thompson \cite{Massey:1}, Romero et al. \cite{Romero:1},
Herrero et al. \cite{Herrero:1}, Benaglia et al. \cite{Benaglia:1}) is positioned within the 1$\sigma$ TeV error
circle. No catalogued X-ray source from the ROSAT all-sky and pointed survey lies within the 2$\sigma$ TeV 
error circle. Our analysis of archival ASCA GIS data yields a 99\% upper limit (2--10 keV) of 
2.0$\times10^{-12}$ erg cm$^{-2}$ s$^{-1}$ (Fig.~\ref{fig:spectrum}). 
Such results may imply that the energy source for particle acceleration is not co-located with the
TeV source, arising instead from the winds of the young/massive stars of Cygnus OB2, either
individually or collectively, or from an alternative source. The former scenario would generally
favour accelerated hadrons interacting with a local, dense gas cloud, giving rise
to $\pi^\circ$-decay TeV emission. The likely hard TeV spectrum can be explained by (energy-dependent) diffusion 
effects, accelerator age, and accelerator to target distance (see eg. Aharonian \cite{Aharonian:4}). 
There is however at present no strong indication from CO and HII surveys (Leung \& Thaddeus \cite{Leung:1}, 
Dame et al. \cite{Dame:1}, Comeron \& Torra \cite{Comeron:1}) for any co-located dense gas cloud, although
see Mart\'{\i} et al. (\cite{Marti:2}) who discuss a nearby HII cloud in the context that follows immediately below. 
A suggested alternative scenario involves a jet-driven termination shock at which accelerated 
electrons produce synchrotron and TeV inverse-Compton (IC) emission (Aharonian \& Atoyan \cite{Atoyan:1}). 
Such a jet could emanate from a nearby microquasar, possibly a class of high energy $\gamma$-ray source
(see eg. Paredes et al. \cite{Paredes:1}).
In fact two nearby sources, 3EG J2033+4118 and also the EGRET source possibly associated with Cyg X-3  
(Mori et al. \cite{Mori:1}) could be GeV indicators of such a microquasar. 
Remarkably, Cyg X-3 appears to have 
a bi-lobal jet (Mart\'{\i} et al. \cite{Marti:2,Marti:1}) well-aligned with the TeV source, the latter which would be $\sim70$pc from Cyg X-3 in absolute terms if it is at the same distance ($>$8.5 kpc). Future X-ray 
observations will be a crucial constraint on the IC emission in this context. 
We interpret the flux from 
3EG J2033+4118 presently as a MeV/GeV upper limit at the TeV COG (Fig.~\ref{fig:spectrum}), and note that the
directly extrapolated energy flux from the TeV source lies about two orders of magnitude below the 3EG J2033+4118 flux at 
overlapping energies.
Finally we note that earlier
claims for a TeV source (Neshpor et al. \cite{Neshpor:1}, at a $\sim$1
Crab flux level) and flaring episodes coincident with a Cyg X-3 radio flare at energies
$>$40 TeV (Merck \cite{Merck:1}, Krawczynski \cite{Kraw:1})  
positionally consistent with our TeV COG have been reported. These results are however in conflict with our
estimates of the flux level and steady nature of the TeV source assuming they all
have the same origin. 
Further observations with the HEGRA CT-System aimed at confirmation and improving our spectral
and source morphology studies are now underway.

\begin{acknowledgements}
The support of the German ministry for Research and
technology BMBF and of the Spanish Research Council CICYT is gratefully
acknowledged.
We thank the Instituto de Astrof\'{\i}sica de Canarias
for the use of the site and for supplying excellent working conditions at
La Palma. We gratefully acknowledge the technical support staff of the
Heidelberg, Kiel, Munich, and Yerevan Institutes. GPR acknowledges receipt of a von Humboldt fellowship.
\end{acknowledgements}



\end{document}